\documentclass[eqsecnum, aps, prd, superscriptaddress, nofootinbib, 10pt, twocolumn, preprintnumbers]{revtex4-2}

\usepackage{amsmath,amssymb}
\usepackage{amsthm}
\usepackage{mathtools}
\usepackage{appendix}
\usepackage{comment}
\usepackage{graphicx}
\usepackage{grffile}
\usepackage{mathrsfs}
\usepackage{bm}
\usepackage{url}
\usepackage{orcidlink}
\usepackage{hyperref}
\usepackage{xcolor}
\usepackage{soul}
\usepackage{algorithm}
\usepackage{algorithmic}
\usepackage{natbib}
\usepackage{physics}
\usepackage{siunitx}
\AtBeginDocument{\RenewCommandCopy\qty\SI}

\DeclareSIUnit[]\solarmass{\text{\ensuremath{\textup{M}_\odot}}}
\usepackage{acronym}
\usepackage{xspace}

\newcommand{\mc}{\mathcal{M}_\mathrm{c}}
\newcommand{\dl}{d_\mathrm{L}}

\newcommand{\etal}{\textit{et al.}}

\begin{document}
\title{Hybrid algorithm combining matched filtering and convolutional neural networks for searching gravitational waves from binary black hole mergers}

\preprint{RESCEU-27/25}

\author{Takahiro S. Yamamoto\,\orcidlink{0000-0002-8181-924X}}
\email{yamamoto.s.takahiro@resceu.s.u-tokyo.ac.jp}
\affiliation{Research Center for the Early Universe (RESCEU), Graduate School of Science, The University of Tokyo, Tokyo 113-0033, Japan}

\author{Kipp Cannon\,\orcidlink{0000-0003-4068-6572}}
\email{kipp@resceu.s.u-tokyo.ac.jp}
\affiliation{Research Center for the Early Universe (RESCEU), Graduate School of Science, The University of Tokyo, Tokyo 113-0033, Japan}

\author{Hayato Motohashi\,\orcidlink{0000-0002-4330-7024}}
\email{motohashi@tmu.ac.jp}
\affiliation{Department of Physics, Tokyo Metropolitan University, 1-1 Minami-Osawa, Hachioji, Tokyo 192-0397, Japan}

\author{Hiroaki W. H. Tahara\,\orcidlink{0000-0002-2225-316X}}
\email{tahara@tmu.ac.jp}
\affiliation{Department of Physics, Tokyo Metropolitan University, 1-1 Minami-Osawa, Hachioji, Tokyo 192-0397, Japan}

\date{\today}
\begin{abstract}
    Efficient searches for gravitational waves from compact binary coalescence are crucial for gravitational wave observations.
    We present a proof-of-concept for a method that utilizes a neural network taking an \textit{SNR map}, a stack of SNR time series calculated by the matched filter, as input and predicting the presence or absence of gravitational waves in observational data.
    We demonstrate our algorithm by applying it to a dataset of gravitational-wave signals from stellar-mass black hole mergers injected into stationary Gaussian noise.
    Our algorithm exhibits comparable performance to the standard matched-filter pipeline and to the machine-learning algorithms that participated in the mock data challenge, MLGWSC-1.
    The demonstration also shows that our algorithm achieves reasonable sensitivity with practical computational resources.
\end{abstract}
\maketitle
\acresetall
\acrodef{GW}{gravitational wave}
\acrodef{CBC}{compact binary coalescence}
\acrodef{BH}{black hole}
\acrodef{BBH}{binary black hole}
\acrodef{LVK}{LIGO-Virgo-KAGRA}
\acrodef{SNR}{signal-to-nose ratio}
\acrodef{PSD}{power spectral density}
\acrodef{ReLU}{Rectified Linear Unit}
\acrodef{FAP}{false alarm probability}
\acrodef{FAR}{false alarm rate}
\acrodef{AUC}{area under the curve}

\section{Introduction}

\ac{LVK} collaboration has detected 218 \ac{GW} events from \ac{CBC} up to the first part of the fourth observation run~\cite{LIGOScientific:2025slb}.
These detections have provided valuable insights into the population of compact binaries~\cite{LIGOScientific:2025pvj}, the cosmic expansion rate~\cite{LIGOScientific:2017adf, LIGOScientific:2025jau}, the nature of gravity~\cite{LIGOScientific:2025obp}, and other related phenomena.
We expect the detection rate to further increase as detector sensitivities improve and as third-generation interferometers come online.

Efficient data analysis algorithms are essential for detecting and analyzing the \ac{GW} signals.
For \ac{CBC} signals, numerical relativity provides precise predictions of the waveforms, and efficient algorithms for generating waveforms from source parameters are well developed (see Sec.~2 of Ref.~\cite{LIGOScientific:2025yae}).
The matched filter is an efficient and sensitive algorithm for finding signals in noisy data when the waveforms of the expected signals are well modeled.
It is known that the matched filter is equivalent to the maximum likelihood detection if the noise is stationary and Gaussian~\cite{Jaranowski:2005hz}.
In this sense, the matched filter is the optimal method to search for the \ac{GW} signal in noisy strain data.
Practically, the detector noise is non-stationary and contaminated by non-Gaussian noise, particularly short, transient noise (glitches).
In practical pipelines, the matched filter is followed by additional veto and ranking procedures that process all candidates generated by the matched filter.
This follow-up stage is computationally expensive and time-consuming, and it often becomes a bottleneck for the rapid analysis required for multi-messenger astronomy, in which electromagnetic observations are conducted in response to triggers from \ac{GW} observations.

Recent studies have demonstrated that deep learning offers an advantage in computational time compared to standard matched filter algorithms~\cite{Cuoco:2024cdk}.
Deep learning~\cite{Goodfellow-et-al-2016} is a framework for analyzing data that are difficult or infeasible to model mathematically.
A neural network, an estimator that imitates biological neurons connecting and communicating information, is employed to analyze data.
It has numerous tunable parameters corresponding to the strength of the connection between the artificial neurons.
With a given dataset, the parameters of the neural network are optimized so that it can analyze unseen or real-world data.
This optimization is called training.
One of the benefits of deep learning is that, after training, it enables rapid data analysis.
In the pioneering work by George~\& Huerta~\cite{George:2016hay}, they showed that the trained neural network can detect a \ac{GW} signal from a \ac{BBH} merger with a detection probability comparable to that of the matched filter, while reducing the computational time by up to \num{e4} times at maximum.

Since this work shows the possibility of deep learning for \ac{GW} data analysis, various algorithms are proposed~\cite{Wang:2019zaj,Gabbard:2017lja,Schafer:2021cml,Schafer:2021fea,Nousi:2022dwh}.
For example, Nousi~\etal~\cite{Nousi:2022dwh} demonstrated that the residual network trained using real data obtained from the third observation run of \ac{LVK} Collaboration surpasses the performance of the PyCBC~\cite{alex_nitz_2024_10473621}, one of the conventional pipelines.
They applied their network to real data and found eight new candidates~\cite{Koloniari:2024kww}.
These works show deep learning is a computationally efficient and highly sensitive method for \ac{GW} data analysis, achieving results comparable to, or even surpassing, those of conventional pipelines. While there are subtle aspects of the performance comparison~\cite{Nagarajan:2025hws,Koloniari:2025cyt}, the potential of deep learning is unequivocally established.

Deep learning for \ac{GW} data analysis is being extensively explored as a versatile tool, targeting various detections including continuous waves~\cite{Joshi:2023hpx,Joshi:2024bkj,Joshi:2025xdz,Cheung:2025lyf,Yamamoto:2020pus,Yamamoto:2022adl,Dreissigacker:2019edy,Dreissigacker:2020xfr}, transient \acp{GW} from core-collapse supernovae~\cite{Astone:2018uge,Iess:2020yqj,Iess:2023quq,Chan:2019fuz,Edwards:2020hmd,Abylkairov:2024hjf,LopezPortilla:2020odz,Sasaoka:2023kte}, and stochastic \ac{GW} background~\cite{Yamamoto:2022kuh,Einsle:2025xsh} (see also Ref.~\cite{Cuoco:2024cdk} as a comprehensive review).
Some algorithms employing deep learning have already been deployed as part of a pipeline analyzing real data (e.g., GWAK~\cite{Raikman:2023ktu}).

The preprocessing, in which the strain data is transformed into input data for the neural network, is a crucial stage that affects the sensitivity of the algorithm.
One of the convenient preprocessing steps is whitening the strain data to suppress the frequency band dominated by detector noise.
Time-frequency maps (e.g., a stack of many short-time Fourier transforms and Q-transforms) are also useful as inputs to neural networks (e.g.,~\cite{Joshi:2023hpx, Sasaoka:2023kte, Sakai:2021rks}).
The matched filter is also useful for the preprocessing stages.
In Refs.~\cite{Beveridge:2023bxa, McLeod:2024pfl,Beveridge:2025wlc}, the \ac{SNR} time series of the best-fit template is fed to the neural network.

Templates with slightly different source parameters are partially correlated.
By incorporating this correlation, the sensitivity of the search algorithm could improve.
This work presents a new idea in which a neural network uses \ac{SNR} time series derived from the matched filter, not only for the best-fit template but also for all templates in the search bank.
We expect that using all \ac{SNR} time series from the template bank will improve the sensitivity of the neural network.
We construct an image by stacking the \ac{SNR} time series and feed it to the neural network to detect \ac{BBH} signals.
This paper presents a proof of principle for this idea.

This paper is organized as follows.
In Sec.~\ref{sec: method}, we explain our algorithm and the dataset used for training the neural network and assess its sensitivity.
After showing the results in Sec.~\ref{sec: result}, we conclude our results and discuss the possible direction of this work in Sec.~\ref{sec: conclusion}.
The code used in this work is available from \url{https://github.com/tsyamamoto21/gw-hybridmfcnn-stellarmassbbh}.

\section{Method}
\label{sec: method}

\begin{figure}[ht]
    \centering
    \includegraphics[width=1.0\linewidth]{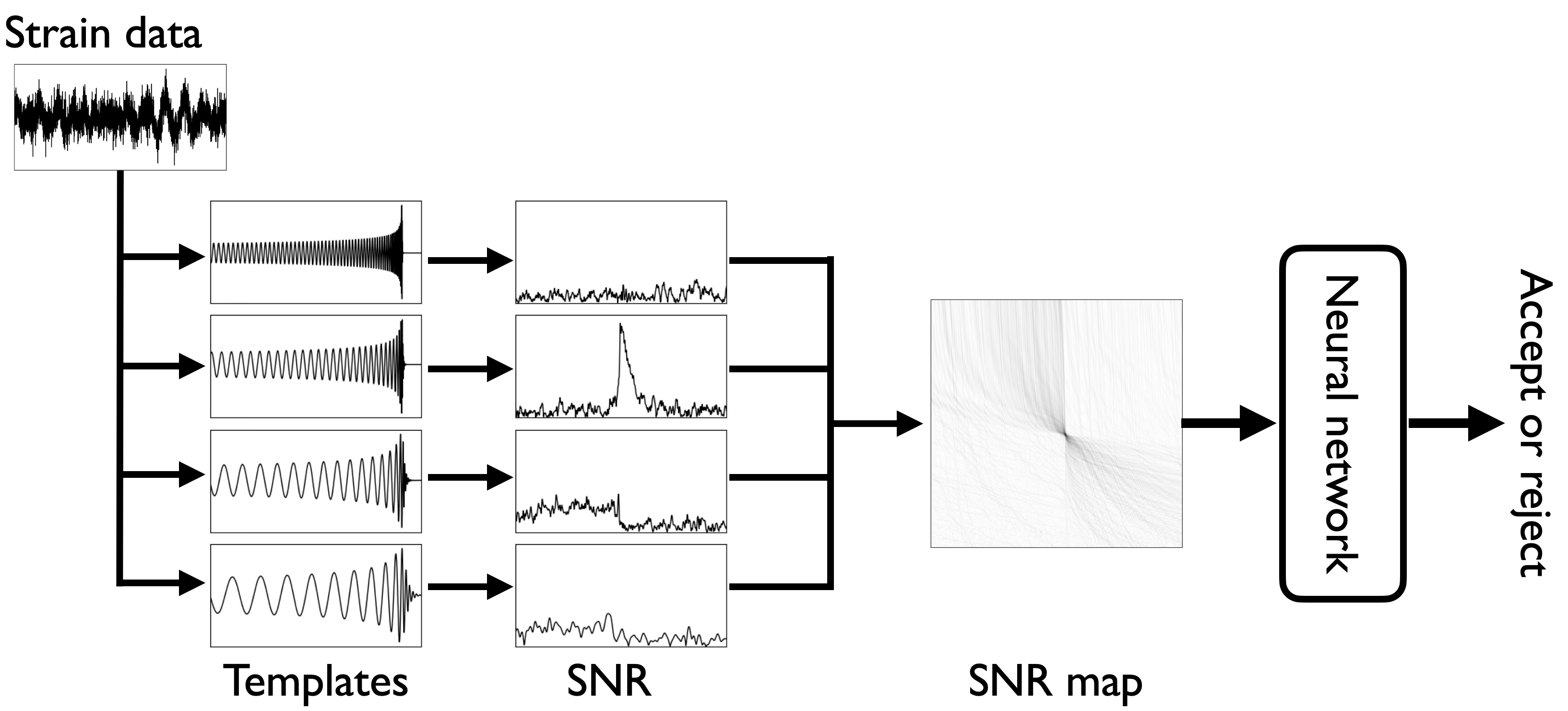}
    \caption{Workflow of our algorithm}
    \label{fig: workflow}
\end{figure}

This section explains the details of our algorithm.
Figure~\ref{fig: workflow} shows the workflow of the algorithm.
First, the matched filter is applied to the observed strain data using a set of waveform templates.
The obtained SNR time series are concatenated into an image (SNR map).
The neural network processes the SNR map and predicts the absence or presence of a \ac{BBH} signal.

\subsection{Matched filter and SNR map}

In this paper, we use the following notation.
The discretized time-domain data is denoted by $x[j] (j=0, 1, \cdots N-1)$.
The discrete Fourier transform of the time domain data is
\begin{equation}
    \tilde{x}[k] = \Delta t \sum_{j=0}^{N-1} x[j] e^{-2\pi ijk/N}\,,
\end{equation}
where $\Delta t$ is the interval of the time samples.
The inverse Fourier transform is
\begin{equation}
    x[j] = \Delta f \sum_{k=0}^{N-1} \tilde{x}[k] e^{2\pi ijk/N}\,,
\end{equation}
where $\Delta f = (N \Delta t)^{-1}$ is the interval of the frequency samples.

The Fourier transform of the strain data is given by
\begin{equation}
    \tilde{d}_I[k] = \tilde{n}_I[k] + \tilde h_{\mathrm{inj},I}[k]\,,
\end{equation}
where $n$ is the detector noise and $h_\mathrm{inj}$ is a \ac{CBC} signal.
The subscript $I$ indicates the interferometers.
The matched filter is a technique to find the predicted signal buried in noisy data.
In this work, we assume that we can accurately simulate the signal expected to be present in the data.
The correlation between the data and the template is a measure of the significance of the signal.

\begin{figure}[ht]
    \centering
    \includegraphics[width=1.0\linewidth]{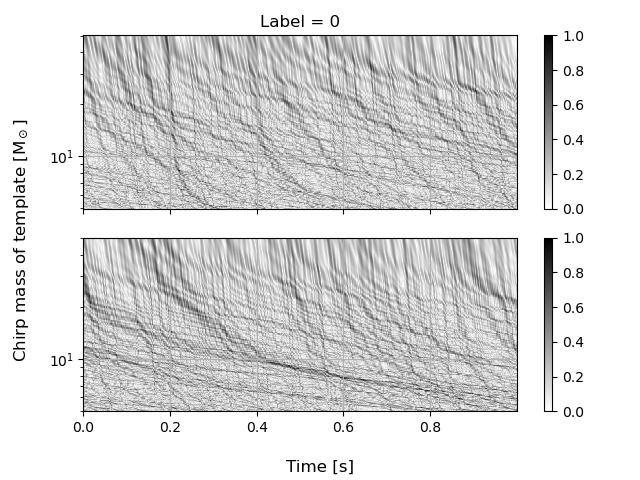}
    \caption{Example of an SNR map for the strain data containing only Gaussian noise. The upper and bottom panels show the data assuming LIGO Hanford and Livingston, respectively. The values are normalized so that the highest and the lowest values must be 1 and 0, respectively (see Eq.~\eqref{eq: normalization snr map}).}
    \label{fig: example noise}
\end{figure}

\begin{figure}[ht]
    \centering
    \includegraphics[width=1.0\linewidth]{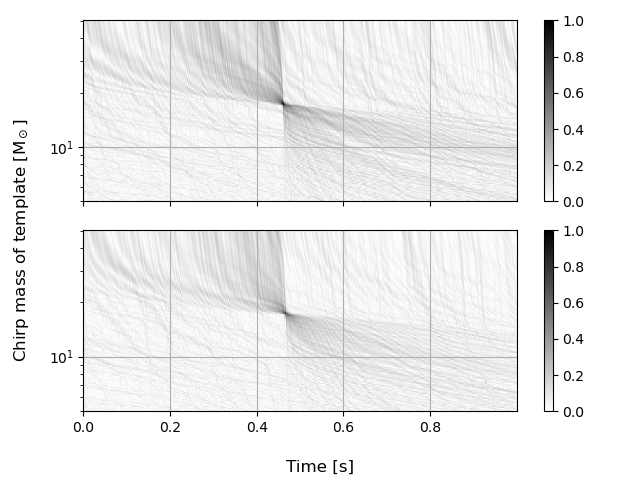}
    \caption{Example of an SNR map for the data containing the simulated Gaussian noise and the injected CBC signal.
    Upper and lower panels correspond to LIGO Hanford and Livingston, respectively.
    Injected signal has the chirp mass of \qty{17.2}{\solarmass}, and its \acp{SNR} are \num{28.4} for Hanford and \num{22.4} for Livingston.
    The normalization~\eqref{eq: normalization snr map} is applied.
    A characteristic pattern of the injected signal is clearly visible around \qty{0.5}{s}.}
    \label{fig: example cbc}
\end{figure}

The noise-weighted inner product between the strain data and the template labeled by an index $r$ is defined by
\begin{equation}
    z_I[j; r] = 4\Delta f \sum_{k=1}^{\lfloor (N-1)/2 \rfloor} \frac{\tilde{d}_I[k] \tilde{h}^\ast[k; r]}{S_I[k]} e^{2\pi ijk/N}\,,
\end{equation}
where ${}^\ast$ is a complex conjugate, and $\lfloor \cdot \rfloor$ is a floor function.
The \ac{PSD} of the detector noise of the $I$ th interferometer is denoted by $S_I$, which is defined by
\begin{equation}
    S_I[k] = \frac{2}{N\Delta t} \langle \abs{\tilde{n}_I[k]}^2 \rangle\,.
\end{equation}
The matched-filter \ac{SNR} is defined by
\begin{equation}
    \rho_I[j; r] = \frac{\abs{z_I[j; r]}}{\sigma_I[r]}\,,
    \label{eq: matched filter snr}
\end{equation}
where $\sigma_I[r]$ is the normalization factor of the template $r$,
\begin{equation}
    \sigma_I[r] = \bqty{ 4\Delta f \sum_{k=1}^{\lfloor (N-1)/2 \rfloor} \frac{\abs{ \tilde{h}[k; r] }^2}{S_I[k]} }^{1/2}\,.
\end{equation}
In our algorithm, we treat $\rho_I[j; r]$ as an image to be input to a neural network and call it an \textit{\ac{SNR} map}.
Figures~\ref{fig: example noise} and \ref{fig: example cbc} show examples of SNR maps for pure-noise data and for a signal injected into noise, respectively.
We used 256 templates with the chirp masses evenly spaced in the logarithmic scale from $\qty{5}{\solarmass}$ to $\qty{50}{\solarmass}$, the common mass ratio $m_1 / m_2 = 1$, and zero spin.
For \qty{1}{s} data with the sampling rate of \qty{2048}{Hz}, an image has the size of (width, height) $= (2048, 256)$.


We may need a large network architecture of the neural network to process an image with a size of (2048, 256).
In general, it is more difficult and often requires more advanced techniques to train neural networks with larger architectures.
It would be beneficial to reduce the input image size to efficiently train our neural networks.
We employ a preprocessing in which we smear an \ac{SNR} map over the time axis by
\begin{equation}
    \rho_{\mathrm{s},I}[J; r] = \Bqty{ \frac{1}{N_\mathrm{s}} \sum_{j=0}^{N_\mathrm{s} - 1} \pqty{ \rho_I[j + JN_\mathrm{s}; r] }^2 }^{1/2}\,,
\end{equation}
where $N_\mathrm{s}$ is the kernel size of the smearing.
Before inputting the \ac{SNR} image to a neural network, we normalize the image so that the value range takes from 0 to 1,
\begin{equation}
    \bar{\rho}_I[J; r] = \frac{\rho_{\mathrm{s},I} [J; r] - \rho_\mathrm{min}}{\rho_\mathrm{max} - \rho_\mathrm{min}}\,,
    \label{eq: normalization snr map}
\end{equation}
where
\begin{equation}
    \rho_\mathrm{max} = \max_{I,J,r} \rho_{\mathrm{s}, I}[J; r]\,,
\end{equation}
and similarly for $\rho_\mathrm{min}$.
Normalizing the input data helps stabilize neural network training. 

There are several techniques for the practical implementation of the matched filter, for example, using a window function to prevent aliasing and accounting for its associated power loss.
We follow Allen~\etal~\cite{Allen:2005fk} for the details of the implementation of the matched filter.
We implemented the matched filter and related signal processing (e.g., calculating matches for many templates simultaneously) using the \texttt{PyTorch} library~\cite{Paszke:2019xhz}.
Using vectorized data and GPU acceleration, we efficiently compute SNR maps.

\subsection{Neural network}

Neural networks are efficient data processors designed in analogy to the biological neurons that connect and process complex information.
The fundamental component of a neural network is the artificial neuron, which takes a multidimensional real-valued vector as input and outputs a real value after applying an affine transformation followed by a nonlinear activation function.
Multiple neurons form a layer. A layer receives a multidimensional real-valued vector from the previous layer and passes a transformed vector to the next layer.
The most basic structure of the neural network, a fully-connected network, consists of successive linear layers and activation functions.

There exist many types of layers, each with their own advantages. 
Among them, we employ convolutional layers.
A convolutional layer contains many filters (or kernels) whose sizes are much smaller than those of the original images or the input data.
The output of the convolutional layer is obtained by convolving the input images (or data) with these filters.
A convolutional layer is often followed by a pooling layer, which compresses the data and reduces its dimensionality.
The data processed by the convolutional layers are characterized by three dimensions: height, width, and the number of channels.
The channels are the generalization of the RGB components of ordinary images.
In convolutional layers, the number of channels corresponds to the number of filters.

\begin{table}[t]
    \centering
    \caption{Structure of the neural network we used in this work.
    When the input and output sizes are shown by three numbers, represented by $(C, W, H)$, $C$ is the channel, $W$ and $H$ are the width and height of the data.
    The first convolutional layer has 2 channels corresponding to data from LIGO Hanford and LIGO Livingston.
    We set the kernel size to 5 for the convolutional layers and to 2 for the max pooling layers.
    All convolutional layers have padding of 1, and all pooling layers have no padding.
    When we train the network, we put the softmax layer (Eq.~\eqref{eq: softmax}) after the last linear layer.}
    \label{tab: neural network structure}
    \begin{ruledtabular}
        \begin{tabular}{lll}
            Layer & Input size & Output size \\ \hline
            Conv2D & (2, 256, 256) & (64, 254, 254) \\
            ReLU & (64, 254, 254) & (64, 254, 254) \\
            Maxpool2D & (64, 254, 254) & (64, 127, 127) \\
            Conv2D & (64, 127, 127) & (64, 125, 125) \\
            ReLU & (64, 125, 125) & (64, 125, 125) \\
            Maxpool2D & (64, 125, 125) & (64, 62, 62) \\
            Conv2D & (64, 62, 62) & (64, 60, 60) \\
            ReLU & (64, 60, 60) & (64, 60, 60) \\
            Maxpool2D & (64, 60, 60) & (64, 30, 30) \\
            Conv2D & (64, 30, 30) & (64, 28, 28) \\
            ReLU & (64, 28, 28) & (64, 28, 28) \\
            Maxpool2D & (64, 28, 28) & (64, 14, 14) \\
            Flatten & (64, 14, 14) & (12544) \\
            Linear & (12544) & (64) \\
            ReLU & (64) & (64) \\
            Linear & (64) & (2)
        \end{tabular}
    \end{ruledtabular}
\end{table}

Our neural network is trained to process SNR maps and identify the presence or absence of \ac{GW} signals.
The structure of the neural network is shown in Table~\ref{tab: neural network structure}.
The network consists of the convolutional block and the fully-connected block.
The convolutional block consists of four successive subcomponents, each followed by a max-pooling layer.
The outputs of the convolutional block are two-dimensional feature maps with multiple channels.
We added a flattening layer to reshape it into a one-dimensional vector before feeding it into the fully-connected block.
The fully-connected block consists of two fully-connected layers.
We use \ac{ReLU} as an activation function.
During training and validation, the softmax layer,
\begin{equation}
    y_i = [\mathrm{Softmax}(\bm{x}) ]_i = \frac{e^{x_i}}{e^{x_0} + e^{x_1}}
    \quad (i = 0, 1)\,,
    \label{eq: softmax}
\end{equation}
is placed at the last layer, although omitted in the Table~\ref{tab: neural network structure}.
Here, $\bm{x}$ is the output vector of the fully-connected block.
The index $i=0~(1)$ corresponds to the absence (presence) of the \ac{GW} signal in the input data.
The output of the softmax layer can be interpreted as the probability of the existence of the \ac{GW} signal.

When we test and apply the trained network,
we use \textit{unbounded softmax replacement}~\cite{Schafer:2021fea} in which we remove the softmax layer and define the detection statistic $\Lambda$ by
\begin{equation}
    \Lambda \coloneqq x_1 - x_0\,.
    \label{eq: detection statistic}
\end{equation}

The structure shown in Table~\ref{tab: neural network structure} is chosen based on the comparison of different model structures and their associated hyperparameters. The comparison is summarized in Appendix~\ref{app: hyperparameter tuning}.

\subsection{Dataset}
\label{subsec: dataset}

We generate four datasets: (1) training dataset, (2) validation dataset, (3) test dataset, and (4) mock data challenge dataset.
The first two are used when we train and optimize our neural network.
The training dataset is used to update the tunable parameters of our neural network, while the loss of the validation dataset is monitored to detect possible overfitting to the training data.
Each of the training dataset and the validation dataset consists of two data pools: the pure-noise images and the pure-signal images.
A pure-noise image is the matched filter with the simulated noise data, no signal is injected.
We generate the Gaussian noise using the \ac{PSD} of \texttt{aLIGOZeroDetHighPower}~\cite{LIGODesign}.
A pure-signal image is the matched filter with the simulated data without noise injection.
At each training iteration, we then create various realizations of \ac{SNR} maps by data augmentation described in the subsection~\ref{subsec: training process} (see also Appendix~\ref{appendix: technique for data augmentation}).

In this work, we focus on the non-spinning \ac{BBH} mergers.
A \ac{CBC} signal is characterized by the intrinsic parameters, component masses in this work, and the extrinsic parameters (e.g., inclination angle).
They are randomly drawn from the distribution shown in Table~\ref{tab: parameter distributions for test data}.
We use the phenomenological waveform \texttt{IMRPhenomXPHM}~\cite{Pratten:2020ceb} for injections and the matched filter templates.

The test dataset is used for assessing the performance of the trained neural network.
It consists of background data, which contain simulated Gaussian noise only, and foreground data, in which a \ac{CBC} signal is injected into simulated Gaussian noise.
The source parameters are randomly drawn from the distributions presented in the Table~\ref{tab: parameter distributions for test data}.
Using the background and foreground data, we draw the curve of the detection probability as a function of the false alarm probability.
The curve enables us to define the \ac{AUC} as an indicator of the neural network's performance.
We compare \ac{AUC} for many sets of hyperparameters in Appendix~\ref{app: hyperparameter tuning}.
We also use the background data to set the threshold value for the detection statistic.

We generate the mock data challenge dataset to compare our algorithm with others that participated in the MLGWSC-1 project~\cite{Schafer:2022dxv}.
We employ dataset 1 of MLGWSC-1 for comparing the algorithms.
The dataset is generated by using publicly available code~\cite{marlin_schafer_2022_7107410}.

\begin{table}[h]
    \caption{Summary of the distribution from which the parameters are drawn for the test dataset.
    The chirp distance $d_\mathrm{c}$ is defined by $d_\mathrm{c} = d_\mathrm{L} (1.4 \cdot 2^{-1/5}\unit{\solarmass} / \mc)^{5/6}$ with the luminosity distance $\dl$.
    The range of the chirp distance is much closer to us compared to that of the MLGWSC-1 dataset.
    This is because we focus on the detectable binaries during the training, validation, and testing.}
    \label{tab: parameter distributions for test data}
    \begin{ruledtabular}
        \begin{tabular}{ll}
            Parameter & Distribution \\ \hline
            Component masses & $\mathcal{U}[10,50]$ \si{\solarmass} \\
            Coalescence phase & $\mathcal{U}[0, 2\pi]$ \\
            Polarization & $\mathcal{U}[0, 2\pi]$ \\
            Inclination & $\cos\iota \sim \mathcal{U}[-1, 1]$ \\
            Declination & $\sin\delta \sim \mathcal{U}[-1, 1]$ \\
            Right ascension & $\mathcal{U}[-\pi, \pi]$ \\
            Chirp-distance & Uniform in volume $[15, 150]$ \si{Mpc}
        \end{tabular}
    \end{ruledtabular}
\end{table}

\subsection{Training process}
\label{subsec: training process}

The loss function to be minimized during the training is the cross-entropy\,,
\begin{equation}
   \ell(\bm{y}, \bm{t}) = - \sum_{i=0,1} t_i \ln y_i\,,
   \label{eq: cross entropy}
\end{equation}
where $\bm{y}$ is an output of the neural network with the softmax layer, and $\bm{t}$ is a label associated with the input data.
The label $\bm{t}$ in Eq.~\eqref{eq: cross entropy} is shown in the one-hot representation, in which $\bm{t} = (1, 0)$ is for the absence of the \ac{GW} signal and $\bm{t} = (0, 1)$ is for the presence of the \ac{GW} signal.

We use the Adam optimizer~\cite{Kingma:2014vow} with the learning rate of \num{1e-4}.
At each training step, the neural network takes a set of input data, which is called a batch, and returns its prediction for each input data.
The sample average of the loss function (Eq.~\eqref{eq: cross entropy}) is calculated and used for the Adam optimization.
The size of a batch is fixed at 128.

When we take a batch for each training step, we first randomly choose a noise \ac{SNR} map and a pure-signal \ac{SNR} map.
With the extrinsic parameters and the \ac{SNR} randomly drawn, we appropriately combine the noise \ac{SNR} map and the pure-signal \ac{SNR} map to create a new \ac{SNR} map.
The detail of this procedure is summarized in Appendix~\ref{appendix: technique for data augmentation}.
This procedure enables us to effectively increase the amount of training data and prevent the neural network from falling into overfitting.

We employ the curriculum training~\cite{George:2016hay} to efficiently train the neural network.
In the early stage of training, the \ac{SNR} of the signal is adjusted to be so large as to be detected easily.
As the training goes, we gradually reduce the signal's \ac{SNR} and, in the end, we adjust the \ac{SNR} within the practically expected range of \acp{SNR}.
We started the training with the \ac{SNR} range from 15 to 40.
The maximum \ac{SNR} was fixed at 40, while the minimum SNR was successively reduced to 10, 8, 6, and 5 at the epochs of 4, 16, 31, and 61, respectively.
We trained the neural network for 150 epochs in total.

\subsection{Processing dataset of MLGWSC-1}
\label{subsec: process mlgwsc-1}

For the MLGWSC-1 dataset, we first segment the data into \qty{16}{s} intervals with an \qty{8}{s} overlap.
The \acp{PSD} are then estimated, and the matched filter is performed on each \qty{16}{s} segment, yielding many SNR maps of \qty{16}{s} widths.
The first and last \qty{4}{s} of the data are trimmed.
These wide SNR maps are divided into SNR maps with \qty{1}{s} widths \qty{0.5}{s} overlaps.
Our algorithm processes these \qty{1}{s} SNR maps by the trained neural network and gets detection statistics.
From Eqs.~\eqref{eq: softmax} and~\eqref{eq: detection statistic}, $\Lambda \geq 0$ is equivalent to the probability of CBC dominating the probability of noise.
In this work, we store only the segments that satisfy $\Lambda \geq 0$ as triggers.
The final outputs of our algorithm are formatted according to the format used in MLGWSC-1 which consists of triplets containing the time of the trigger, the value of the detection statistic, and the variance of the trigger time.
The time of the trigger is the time at which a \ac{CBC} signal is predicted to exist by the algorithm.
The variance of the trigger time represents a window centered on the time of the trigger, indicating that a signal is expected to exist within that window.
We use the time at the center of the segment as the time of the trigger.
We fix the variance of the trigger time at \qty{0.25}{s} because our neural network is trained to detect the signal that is located within the \qty{0.5}{s} window located at the center of the SNR map.
After storing all trigger candidates, they are sorted in the order of the time of the trigger.
Two or more trigger candidates overlapping each other accounting their variances are merged as a single trigger.
The new trigger has the trigger time at the center of the merged candidates.
The detection statistic of the new trigger is the maximum value of the detection statistics among the merged candidates.
The variance of the trigger time is set to half of the total window length of the merged candidates.
We repeat concatenating triggers until all triggers do not overlap each other.

The sensitive volume as a function of the false alarm rate are used as metrics of the pipeline's sensitivity.
The false alarm rate is calculated by counting the false alarms and dividing it by the duration of the background data,
\begin{equation}
    \mathcal{F}(\Lambda_\mathrm{th}) = \frac{N_\mathrm{bg} (\Lambda_\mathrm{th})}{T}\,,
    \label{eq: far}
\end{equation}
where $N_\mathrm{bg}(\Lambda_\mathrm{th})$ is the number of triggers in the background data that exceed the threshold $\Lambda_\mathrm{th}$.
We also count the number of triggers that exceed the threshold $\Lambda_\mathrm{th}$ in the foreground data.
Using Eq.~\eqref{eq: far}, we describe the number of triggers as a function of the false alarm rate $\mathcal{F}$ and denote it by $N_\mathrm{fg}(\mathcal{F})$.
The sensitive volume can be calculated by
\begin{equation}
    V(\mathcal{F}) \simeq \mathrm{Vol}(d_\mathrm{max}) \frac{N_\mathrm{fg}(\mathcal{F})}{N_\mathrm{inj}}\,,
\end{equation}
where $N_\mathrm{inj}$ is the total number of injections, and $\mathrm{Vol}(d)$ is the volume of the three-dimensional sphere with the radius of $d$.
The false alarm rate and the sensitive volume are evaluated by the script provided by MLGWSC-1 project~\cite{Schafer:2022dxv}.

\section{Result}
\label{sec: result}

\subsection{Training and validation}

\begin{figure}
    \centering
    \includegraphics[width=1.0\linewidth]{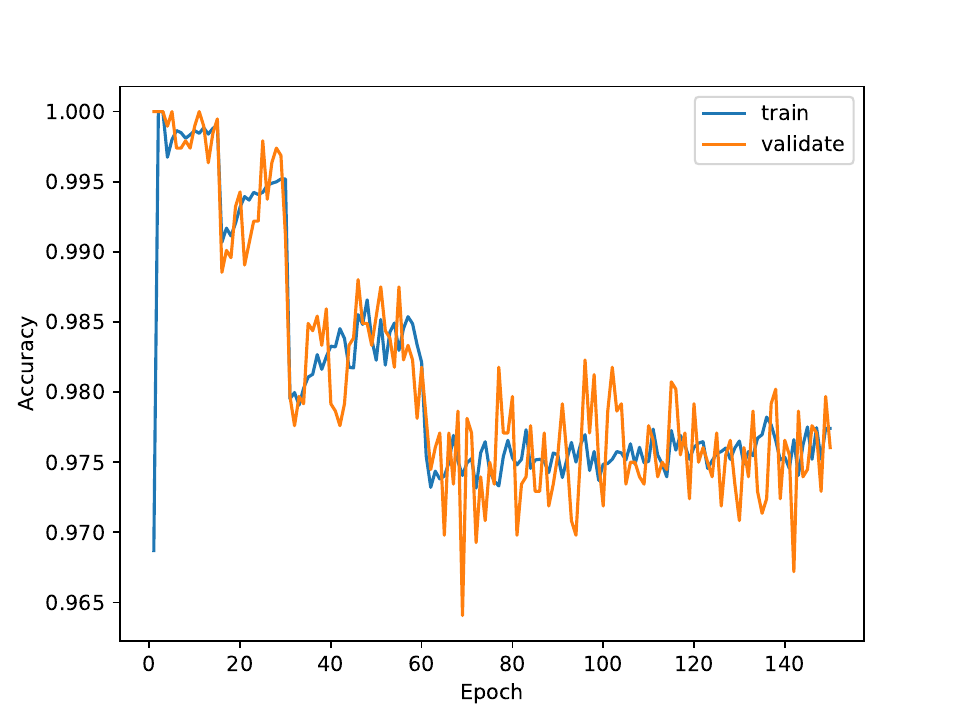}
    \caption{Accuracy curve. The blue and orange lines represent the accuracy for the training data and validation data, respectively.
    The sudden decreases at the epochs of 15, 30, and 60 are when the \ac{SNR} range of the training data is changed to include the low \ac{SNR} data.}
    \label{fig: accuracy curve}
\end{figure}

Figure~\ref{fig: accuracy curve} shows the accuracy for the training data and the validation data during the training.
There are sudden drops in the accuracy at some epochs.
Those epochs correspond to those at which the \ac{SNR} range of the training data has changed to include low \ac{SNR} data in the training and the validation data.
The accuracies of the training data and the validation data appear to have a different amplitude of the fluctuation.
This is likely due to the difference in the number of data points used per epoch for the training and validation datasets.
Besides this difference in the amplitude of the fluctuation, the validation accuracy follows the training accuracy; the neural network does not overfit.

\begin{figure}
    \centering
    \includegraphics[width=1.0\linewidth]{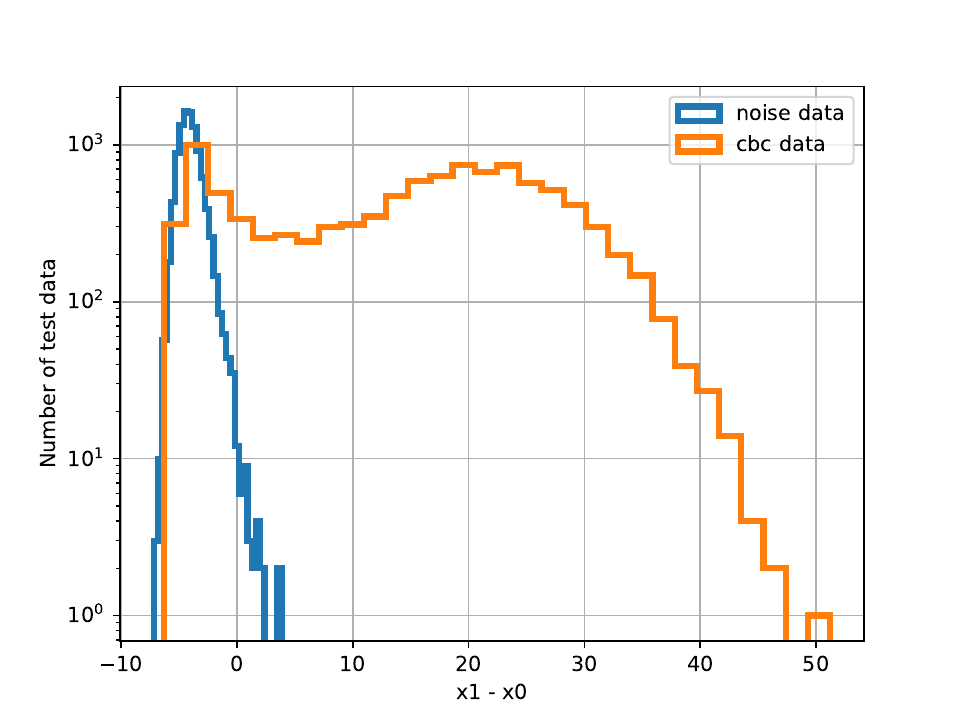}
    \caption{Distributions of the statistics for the noise data set and the CBC datasets. The blue and the orange lines are the results for the noise data and the CBC data, respectively.}
    \label{fig: hist usr cbc}
\end{figure}

Figure~\ref{fig: hist usr cbc} shows the histogram of the detection statistics for the test datasets.
The noise data exhibits a single peak, while the \ac{CBC} data shows a bimodal distribution, with the larger peak corresponding to the observable signal and the smaller peak corresponding to data misclassified as noise by the neural network.

\subsection{Comparison with algorithms of MLGWSC-1}

\begin{figure*}
    \centering
    \includegraphics[width=1.0\linewidth]{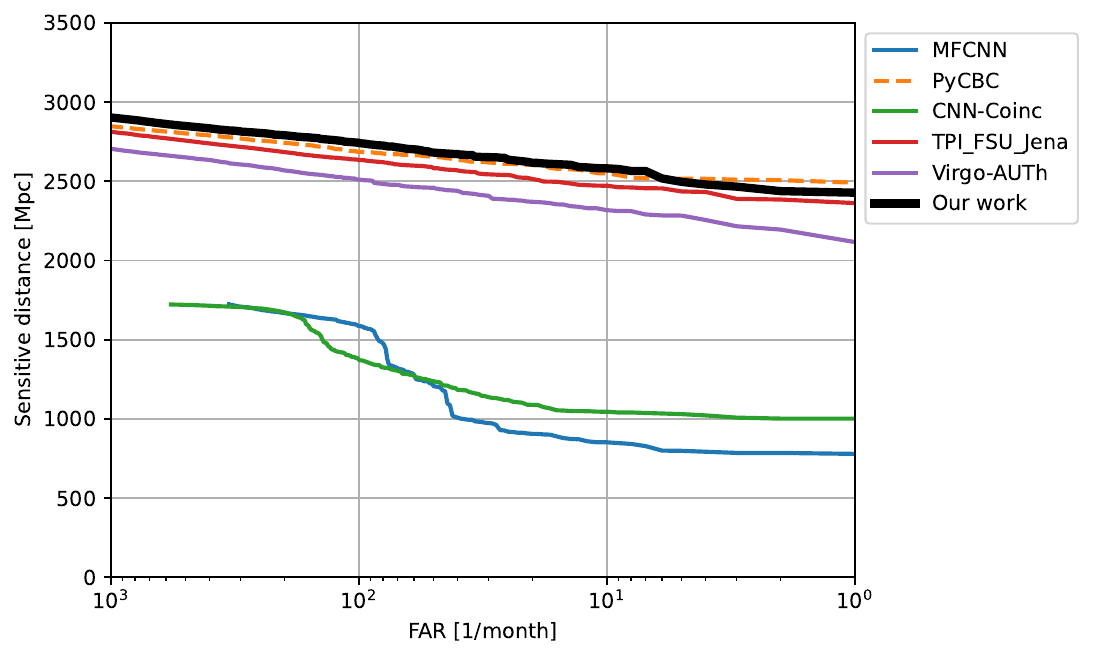}
    \caption{Sensitivity of each method as a function of the false alarm rate. The black solid line is our method. The colored lines show the results of MLGWSC-1, which is publicly available from~\cite{marlin_schafer_2022_7107410}. The colored solid lines are the machine learning methods, and the colored dashed line is the PyCBC, which is a standard matched filter method.}
    \label{fig: sensitivity comparison}
\end{figure*}

\begin{table*}[t]
    \centering
    \caption{Sensitivities and runtimes of our method and the methods that participated in the MLGWSC-1.
    All numbers except our method are taken from Table~IV of~\cite{Schafer:2022dxv}.
    Runtimes are presented to show that our method is feasible within a realistic computational time, not to say our method is more efficient than the others.
    The standard hardware used in the MLGWSC-1 is $2\times$ Intel Xeon Silver 4215 8(16) cores(threads) at \qty{2.5}{GHz} and $8\times$ NVIDIA RTX 2070 Super (8GB VRAM).
    `$\ast$' implies that the time is approximated since PyCBC did not use the computer cluster for GWSCML-1.
    `$\dagger$' shows that our method uses a different hardware architecture from those used for MLGWSC-1.
    Our method uses a CPU of Intel(R) Xeon(R) Platinum 8468 (\qty{2.1}{GHz} / \qty{48}{core}) and a GPU NVIDIA H100.}
    \label{tab: mlgwsc-1 results}
    \begin{ruledtabular}
        \begin{tabular}{lllllll}
            Method & \multicolumn{3}{c}{Sensitivity [Mpc] at FAR = $x$ per month} & \multicolumn{3}{c}{Runtime [s]} \\
             & $x=100$ & $x=10$ & $x=1$ & foreground & background & average \\ \hline
            MFCNN & 1586.90 & 852.18 & 779.21 & 42842 & 43820 & 43331 \\
            PyCBC & 2686.55 & 2550.57 & 2491.53 & $5406^\ast$ & $5092^\ast$ & $5249^\ast$ \\
            CNN-Coinc & 1372.30 & 1045.34 & 1001.55 & 14003 & 12996 & 13500 \\
            TPI FSU Jena & 2634.80 & 2472.31 & 2362.51 & 3758 & 3530 & 3644 \\
            Virgo-AUTh & 2511.95 & 2317.53 & 2116.38 & 5490 & 5520 & 5505 \\
            Our method & 2743.22 & 2581.85 & 2428.10 & $2802^\dagger$ & $2834^\dagger$ & $2819^\dagger$
        \end{tabular}
    \end{ruledtabular}
\end{table*}

We applied the trained network to the dataset 1 of the MLGWSC-1.
Our method identified \num{31060} triggers in the background data and \num{62060} triggers in the foreground data.
The results are summarized in Fig.~\ref{fig: sensitivity comparison} and Table~\ref{tab: mlgwsc-1 results}.
Our algorithm shows the sensitivity distance of \qty{2428.10}{Mpc} at \ac{FAR} of 1 per month, which is comparable to those of PyCBC and other machine learning methods.
The sensitivity changes only mildly as the \ac{FAR} varies.
Since the hardware architectures are not common for all methods, the reader should not interpret the runtime results as indicating that our method is necessarily more efficient than the others.
We conclude that our algorithm achieves good sensitivity and is practically applicable in terms of computational time.

\section{Conclusion}
\label{sec: conclusion}

We proposed a new algorithm that combines matched filter and deep learning.
Our neural network uses the SNR time series of the matched filter as an input image and identifies \ac{CBC} signals.
We compared the performance of our method to those of the other algorithms that participated in MLGWSC-1.
Our method shows comparable sensitivity to those of a standard \ac{CBC} detection pipeline and the machine learning methods.
We have shown that our method is practically feasible in terms of sensitivity and computational time.

This work presents a proof of concept of our algorithm.
Several directions exist in which our approach could offer significant advantages for future development.
First, our demonstration focuses on the data in which the \ac{CBC} signals are injected into the simulated Gaussian noise.
However, it is well known that the detector noise is neither perfectly Gaussian nor stationary.
For practical use, we need to train our network using data where \ac{CBC} signals are injected into observational noise samples obtained from publicly available data.

Second, we can extend our approach to high-mass \ac{BBH} systems.
Due to the short duration of the high mass \ac{BBH} in the ground-based detector frequency band, the confusion with glitch noise, a burst-like transient disturbance, is significant, leading to an increase in the false alarm rate.
By integrating the matched filter with a neural network selection, our algorithm can effectively mitigate the impact of glitch noise in this critical mass regime.

Third, we can move to the low-mass \ac{CBC} signals, such as binary neutron stars and neutron star-black hole binaries.
Such systems typically have masses lower than \acp{BBH}, and \acp{GW} from those systems have long duration.
Therefore, the required number of templates for the matched filter is significantly larger than that of the \ac{BBH} signals.
In particular, neutron star-black hole binaries can be affected by orbital precession due to misaligned spins, which modifies the waveform.
Furthermore, the asymmetry in the mass of the neutron star-black hole binary can efficiently excite higher modes such as $(\ell, m) = (3,3)$ and $(4,4)$.
The template bank with the aligned spin and the dominant modes, which is used by the existing pipelines, can lose a few tens percent of the \ac{SNR} due to the template bank mismatch~\cite{Dhurkunde:2022aek}.
We can consider using waveform models that incorporate these effects, although this requires an increase in the number of templates for matched-filter algorithms.
The computational cost may become impractical for low-latency and online analysis.

Finally, we can consider incorporating our algorithm into existing pipelines.
The current \ac{CBC} searches usually use a low threshold of the detection statistic not to miss any astrophysical signals.
This choice, however, increases the expected number of false alarms.
The current \ac{CBC} search pipeline takes a lot of time to screen the numerous candidates that are returned by the matched filter.
Our approach provides an efficient mechanism for candidate screening, significantly reducing the overall computational time required for the search.

\acknowledgments

We used the software \texttt{numpy}~\cite{Harris:2020xlr}, \texttt{scipy}~\cite{Virtanen:2019joe}, \texttt{matplotlib}~\cite{Hunter:2007ouj}, \texttt{astropy}~\cite{Astropy:2013muo, Astropy:2018wqo, Astropy:2022ucr}, \texttt{PyCBC}~\cite{alex_nitz_2024_10473621}, \texttt{PyTorch}~\cite{Paszke:2019xhz}.
This research used computational resources of Pegasus provided by Multidisciplinary Cooperative Research Program in Center for Computational Sciences, University of Tsukuba.
This work was supported in part by JSPS KAKENHI Grant No.~JP23K13099, JP23H04502 (T.S.Y), JP23H04893 (K.C.), and JP22K03639 (H.M.).

\appendix
\section{Techniques for data augmentation}
\label{appendix: technique for data augmentation}

The performance of a trained neural network critically depends on the volume and variability of the training dataset.
When the dataset volume is limited, data augmentation is a useful technique to effectively increase the number of training samples and enhance the generalization.
In this work, the neural network utilizes SNR maps computed by matched filter as input data.
Because the matched filter is computationally expensive, we cannot generate datasets with enough volume and variability.
Particularly, we may have only a limited number of combinations of noise data and injected signals.
Relying on such limited training samples, the neural networks will tend to overfit.
To mitigate this risk of overfitting, we employ a strategy in which the training data are dynamically generated, based on pools of the noise data and the \ac{GW} signal data.
For each iteration step, we choose a combination of noise data and signal data and combine them appropriately so that the resulting SNR map matches a randomly sampled target SNR value.
In this appendix, we explain the mathematical framework and detailed description underlying the data augmentation procedure.

It is convenient to redefine the matched filter template $\tilde{h}^\ast[k; r]$ by dividing the normalization factor $\sigma[r]$, i.e.,
\begin{equation}
    \frac{\tilde{h}^\ast[k; r]}{\sigma[r]} \to \tilde{h}^\ast[k; r]\,.
\end{equation}
With this notation, the \ac{SNR} (Eq.~\eqref{eq: matched filter snr}) takes the form
\begin{align}
    \rho_I[j; r] &= \abs{ z_I[j; r] } \notag \\
    &= \abs{ 4\Delta f \sum_{k=1}^{\lfloor (N-1)/2 \rfloor} \frac{\tilde{d}_I[k] \tilde{h}^\ast[k; r]}{S_I[k]} e^{2\pi ijk/N} }\,.
\end{align}

We can decompose $z_I$ into
\begin{equation}
    z_I = z_{\mathrm{n}, I} + z_{\mathrm{inj}, I}\,,
    \label{eq: z = zinj + zn}
\end{equation}
where
\begin{equation}
    z_{\mathrm{n},I}[j; r]
    = 4\Delta f \sum_{k=1}^{\lfloor (N-1)/2 \rfloor} \frac{\tilde{n}_I[k] \tilde{h}^\ast[k; r]}{S_I[k]} e^{2\pi ijk/N}\,,
\end{equation}
and
\begin{equation}
    z_{\mathrm{inj},I}[j; r]
    = 4\Delta f \sum_{k=1}^{\lfloor (N-1)/2 \rfloor} \frac{\tilde{h}_{\mathrm{inj},I}[k] \tilde{h}^\ast[k; r]}{S_I[k]} e^{2\pi ijk/N}\,.
    \label{eq: z_inj}
\end{equation}

A \ac{CBC} signal is characterized by several parameters that can be classified into the intrinsic parameters and the extrinsic parameters.
The intrinsic parameters represent the source parameters, such as the chirp mass and component spins.
The extrinsic parameters describe the geometrical information of the source and the detector.
The source direction is denoted by the right ascension $\alpha$ and the declination $\delta$. The polarization and the inclination angle are denoted by $\psi$ and $\iota$, respectively.
The injected signal is
\begin{align}
    \tilde{h}_{\mathrm{inj},I} = \mathcal{A} &\bigg\{ F_{+,I}(\alpha, \delta, \psi) A_+(\iota) h_\mathrm{inj}^\mathrm{p}[k] \notag \\
    &\quad + F_{\times, I} (\alpha, \delta, \psi) A_\times(\iota) h_\mathrm{inj}^\mathrm{c}[k] \bigg\}\,.
\end{align}
where $F_{+,I}$ and $F_{\times, I}$ are the antenna pattern functions of $I$ the detector, and we define $A_+ = (1 + \cos^2\iota)/2$ and $A_\times = \cos\iota$.
The plus mode and cross mode waveforms, $h^\mathrm{p}_\mathrm{inj}$ and $h^\mathrm{c}_\mathrm{inj}$, depend only on the intrinsic parameters.
The luminosity distance of the source and the numerical factor are included in an amplitude parameter $\mathcal{A}$.
We define
\begin{equation}
    z_{\mathrm{p/c}, I}[j; r] = 4\Delta f \sum_{k=1}^{\lfloor (N-1)/2 \rfloor} \frac{\tilde{h}_\mathrm{inj}^{\mathrm{p/c}}[k] \tilde{h}^\ast[k; r]}{S_I[k]} e^{2\pi ijk/N}\,.
\end{equation}
Assuming that two interferometers have the same \acp{PSD}, we omit the subscript $I$ from the \ac{PSD} and the $z_\mathrm{p/c}$.
We can reconstruct $z_\mathrm{inj}$ (Eq.~\eqref{eq: z_inj}) by taking a linear combination of $z_\mathrm{p}$ and $z_\mathrm{c}$ as
\begin{equation}
    z_{\mathrm{inj}, I} = \mathcal{A} \pqty{ F_{+, I} A_+ z_\mathrm{p} + F_{\times, I} A_\times z_\mathrm{c} }
    =: \mathcal{A} \tilde{z}_{\mathrm{inj}, I}\,,
    \label{eq: zinj and normalization factor}
\end{equation}
where the amplitude parameter $\mathcal{A}$ will be chosen so that the total \ac{SNR} equals the target value.

The pure-noise pool contains 1250 (125) realizations of $z_{\mathrm{n}, I}$ for the training (validation).
Each of them has a duration of \qty{8}{s}.
We randomly crop a part of \qty{1}{s} of the pure-noise SNR map.
In this work, we assume LIGO's two detectors have the same sensitivity.
Therefore, we ignore the detector dependence in $z_{\mathrm{n}, I}$ and use a common pool for the two LIGO detectors.
The pure-signal pool consists of 10000 (1000) realizations of $z_\mathrm{p}$ and $z_\mathrm{c}$ for training (validation), and they have a duration of \qty{1}{s}.
For each iteration step, we randomly draw the samples of extrinsic parameters and the total \ac{SNR}, which we denote by $\varrho_\mathrm{target}$.
We also randomly choose an SNR map $z_{\mathrm{n}}$ of pure noise and
SNR maps ($z_{\mathrm{p}, I}$ and $z_{\mathrm{c}, I}$) of pure signals.
The amplitude parameter $\mathcal{A}$ can be adjusted by using the extrinsic parameters, the total SNR, and the SNR maps as follows.
For each $I$, we get the maximum value of $\abs{ \tilde{z}_{\mathrm{inj}, I} }$ and the corresponding time index $\hat{j}$ and template $\hat{r}$, i.e.,
\begin{equation}
    \abs{ \tilde{z}_{\mathrm{inj}, I} [\hat{j}; \hat{r}] }
    = \max_{j, r} \abs{ \tilde{z}_{\mathrm{inj}, I} [j; r] }\,.
\end{equation}
Defining 
\begin{equation}
    \mathcal{Z}_{\mathrm{inj},I} \coloneqq \tilde{z}_{\mathrm{inj}, I}[\hat{j}; \hat{r}]\,,
\end{equation}
and
\begin{equation}
    \mathcal{Z}_{\mathrm{n}, I} \coloneqq z_{\mathrm{n}, I}[\hat{j}; \hat{r}]\,,
\end{equation}
we can write the \ac{SNR} of the injected signal with $I$ th detector by
\begin{equation}
    [\varrho_I(\mathcal{A})]^2 = \abs{ \mathcal{Z}_{\mathrm{inj}, I} + \mathcal{A} \mathcal{Z}_{\mathrm{n}, I} }^2\,.
    \label{eq: varrho sq I}
\end{equation}
The total \ac{SNR} $\varrho_\mathrm{tot}$ is defined by
\begin{equation}
    \varrho_\mathrm{tot}(\mathcal{A}) = \sqrt{\sum_{I=1,2} [\varrho_I(\mathcal{A})]^2}\,.
    \label{eq: varrho tot}
\end{equation}
Substituting Eq.~\eqref{eq: varrho sq I} into Eq.~\eqref{eq: varrho tot}, we get
\begin{equation}
    [\varrho_\mathrm{tot}(\mathcal{A})]^2
    = a\mathcal{A}^2 + b\mathcal{A} + c\,,
\end{equation}
where
\begin{align}
    a &= \sum_{I=1,2} \abs{ \mathcal{Z}_{\mathrm{inj}, I}}^2\,, \\
    b &= 2 \Re{ \sum_{I=1,2} \mathcal{Z}_{\mathrm{n}, I} \mathcal{Z}^\ast_{\mathrm{inj}, I} }\,,\\
    c &= \sum_{I=1,2} \abs{\mathcal{Z}_{\mathrm{n}, I}}^2\,.
\end{align}
Solving the equation $[\varrho_\mathrm{tot} (\mathcal{A})]^2 = \varrho_\mathrm{target}^2$, we get the appropriate amplitude factor $\mathcal{A}$.

Depending on the source position, we also take into account the difference in arrival time between two detectors.
We shift the two \ac{SNR} maps to account for the difference in arrival time.
Finally, we get an \ac{SNR} map consistent with the randomly sampled target \ac{SNR} and extrinsic parameters.

\section{Hyperparameter tuning}
\label{app: hyperparameter tuning}
To choose the architecture of the neural network, it is important to train neural networks with various settings and compare their performances.
This procedure is called hyperparameter tuning.
We tested 13 different settings, such as neural network structure, and compared their performance.
The metric of the performance is the \ac{AUC} on the test data.
Table~\ref{tab: hyperparameters} shows the list of the neural network settings we tried.
Some models experience a situation in which the loss function suddenly returns \texttt{NaN} and the accuracy drops to 0.5.
For models that are trained without such anomalous behavior, the \acp{AUC} are almost comparable.
Here, we choose model No., which shows the highest \ac{AUC} value.

\begin{table*}[t]
    \caption{Hyperparameter sets we tested. The model name `CNN $(C, W, H)$' means that the convolutional layers have $C$ channels, and their kernel size is $(W, H)$.
    All CNN models have four convolutional layers, each of which is followed by a max pooling layer with a kernel size of two.
    The activation functions of ResNet~\cite{He:2015wrn} and ViT~\cite{Dosovitskiy:2020qjv} have already been chosen by their construction.
    The fifth column shows the gradient max norm, which limits the gradient during back propagation and prevents the weights from diverging.
    `CT' is short for curriculum training.
    For some models, training suddenly broke down, leading to, e.g., the calculated loss function showing \texttt{NaN} and/or the measured accuracies suddenly dropping to 0.5.
    The seventh column shows whether the model is trained without any such anomalous behaviors. }
    \label{tab: hyperparameters}
    \begin{ruledtabular}
        \begin{tabular}{llllllll}
            No & model & activation function & smearing kernel & gmn & CT & Stable? & AUC \\ \hline
            1 & CNN (64, 5, 5) & ReLU & 8 & None & $\checkmark$ & Yes & 0.9444 \\
            2 & CNN (64, 13, 13) & ReLU & None & 0.5 & $\checkmark$ & No & - \\
            3 & CNN (64, 5, 13) & LeakyReLU & None & 0.5 & $\checkmark$ & Yes & 0.9374 \\
            4 & CNN (64, 13, 13) & LeakyReLU & None & 0.5 & $\checkmark$ & Yes & 0.9345 \\
            5 & CNN (64, 5, 13) & ReLU & None & 10.0 & $\checkmark$ & Yes & 0.9361 \\
            6 & CNN (64, 5, 13) & ReLU & None & 10.0 &  & Yes & 0.9357 \\
            7 & CNN (256, 5, 5) & ReLU & 8 & None & $\checkmark$ & Yes & 0.9437 \\
            8 & CNN (64, 5, 5) & ReLU & 8 & None &  & No & - \\
            9 & ResNet18 & - & 8 & None & $\checkmark$ & Yes & 0.9431 \\
            10 & ResNet18 & - & None & None & $\checkmark$ & Yes & 0.9367 \\
            11 & ResNet50 & - & 8 & None & $\checkmark$ & Yes & 0.9430 \\
            12 & ViT\_b\_16 & - & 8 & None & $\checkmark$ & Yes & 0.9342 \\
            13 & CNN (64, 5, 5) (zero padding) & ReLU & 8 & None & $\checkmark$ & No & - 
        \end{tabular}
    \end{ruledtabular}
\end{table*}

\bibliography{reference}

@book{Goodfellow-et-al-2016,
    title={Deep Learning},
    author={Ian Goodfellow and Yoshua Bengio and Aaron Courville},
    publisher={MIT Press},
    note={\url{http://www.deeplearningbook.org}},
    year={2016}
}

@article{He:2015wrn,
    author = "He, Kaiming and Zhang, Xiangyu and Ren, Shaoqing and Sun, Jian",
    title = "{Deep Residual Learning for Image Recognition}",
    eprint = "1512.03385",
    archivePrefix = "arXiv",
    primaryClass = "cs.CV",
    doi = "10.1109/CVPR.2016.90",
    month = "12",
    year = "2015",
    journal = ""
}

@article{Dosovitskiy:2020qjv,
    author = "Dosovitskiy, Alexey and others",
    title = "{An Image is Worth 16x16 Words: Transformers for Image Recognition at Scale}",
    eprint = "2010.11929",
    archivePrefix = "arXiv",
    primaryClass = "cs.CV",
    month = "10",
    year = "2020",
    journal = ""
}

@article{LIGOScientific:2017adf,
    author = "Abbott, B. P. and others",
    collaboration = "LIGO Scientific, Virgo, 1M2H, Dark Energy Camera GW-E, DES, DLT40, Las Cumbres Observatory, VINROUGE, MASTER",
    title = "{A gravitational-wave standard siren measurement of the Hubble constant}",
    eprint = "1710.05835",
    archivePrefix = "arXiv",
    primaryClass = "astro-ph.CO",
    reportNumber = "LIGO-P1700296, FERMILAB-PUB-17-472-A-AE",
    doi = "10.1038/nature24471",
    journal = "Nature",
    volume = "551",
    number = "7678",
    pages = "85--88",
    year = "2017"
}

@article{LIGOScientific:2025jau,
    author = "Abac, A. G. and others",
    collaboration = "LIGO Scientific, VIRGO, KAGRA",
    title = "{GWTC-4.0: Constraints on the Cosmic Expansion Rate and Modified Gravitational-wave Propagation}",
    eprint = "2509.04348",
    archivePrefix = "arXiv",
    primaryClass = "astro-ph.CO",
    reportNumber = "LIGO-P2400152",
    month = "9",
    year = "2025",
    journal = ""
}

@article{LIGOScientific:2025yae,
    author = "Abac, A. G. and others",
    collaboration = "LIGO Scientific, VIRGO, KAGRA",
    title = "{GWTC-4.0: Methods for Identifying and Characterizing Gravitational-wave Transients}",
    eprint = "2508.18081",
    archivePrefix = "arXiv",
    primaryClass = "gr-qc",
    reportNumber = "LIGO-P2400300",
    month = "8",
    year = "2025",
    journal = ""
}

@article{LIGOScientific:2025pvj,
    author = "Abac, A. G. and others",
    collaboration = "LIGO Scientific, VIRGO, KAGRA",
    title = "{GWTC-4.0: Population Properties of Merging Compact Binaries}",
    eprint = "2508.18083",
    archivePrefix = "arXiv",
    primaryClass = "astro-ph.HE",
    reportNumber = "LIGO-P2400004",
    month = "8",
    year = "2025",
    journal = ""
}

@article{LIGOScientific:2025obp,
    collaboration = "LIGO Scientific, VIRGO, KAGRA",
    title = "{Black Hole Spectroscopy and Tests of General Relativity with GW250114}",
    eprint = "2509.08099",
    archivePrefix = "arXiv",
    primaryClass = "gr-qc",
    reportNumber = "LIGO P2500461",
    month = "9",
    year = "2025",
    journal = ""
}

@article{Jaranowski:2005hz,
    author = "Jaranowski, Piotr and Krolak, Andrzej",
    title = "{Gravitational-Wave Data Analysis. Formalism and Sample Applications: The Gaussian Case}",
    eprint = "0711.1115",
    archivePrefix = "arXiv",
    primaryClass = "gr-qc",
    doi = "10.12942/lrr-2012-4",
    journal = "Living Rev. Rel.",
    volume = "8",
    pages = "3",
    year = "2005"
}

@article{Dhurkunde:2022aek,
    author = "Dhurkunde, Rahul and Nitz, Alexander H.",
    title = "{Sensitivity of spin-aligned searches for neutron star-black hole systems using future detectors}",
    eprint = "2207.14645",
    archivePrefix = "arXiv",
    primaryClass = "astro-ph.IM",
    doi = "10.1103/PhysRevD.106.103035",
    journal = "Phys. Rev. D",
    volume = "106",
    number = "10",
    pages = "103035",
    year = "2022"
}

@article{Beveridge:2023bxa,
    author = "Beveridge, Damon and McLeod, Alistair and Wen, Linqing and Wicenec, Andreas",
    title = "{Novel deep learning approach to detecting binary black hole mergers}",
    eprint = "2308.08429",
    archivePrefix = "arXiv",
    primaryClass = "gr-qc",
    doi = "10.1103/PhysRevD.111.024005",
    journal = "Phys. Rev. D",
    volume = "111",
    number = "2",
    pages = "024005",
    year = "2025"
}

@article{McLeod:2024pfl,
    author = "McLeod, Alistair and Beveridge, Damon and Wen, Linqing and Wicenec, Andreas",
    title = "{Binary neutron star merger search pipeline powered by deep learning}",
    eprint = "2409.06266",
    archivePrefix = "arXiv",
    primaryClass = "gr-qc",
    doi = "10.1103/PhysRevD.111.024035",
    journal = "Phys. Rev. D",
    volume = "111",
    number = "2",
    pages = "024035",
    year = "2025"
}

@article{Beveridge:2025wlc,
    author = "Beveridge, Damon and McLeod, Alistair and Wen, Linqing and Guo, Weichangfeng and Wicenec, Andreas",
    title = "{Searching for binary black hole mergers with deep learning in Advanced LIGO's third observing run}",
    eprint = "2512.04516",
    archivePrefix = "arXiv",
    primaryClass = "gr-qc",
    month = "12",
    year = "2025",
    journal = ""
}

@article{Allen:2005fk,
    author = "Allen, Bruce and Anderson, Warren G. and Brady, Patrick R. and Brown, Duncan A. and Creighton, Jolien D. E.",
    title = "{FINDCHIRP: An Algorithm for detection of gravitational waves from inspiraling compact binaries}",
    eprint = "gr-qc/0509116",
    archivePrefix = "arXiv",
    doi = "10.1103/PhysRevD.85.122006",
    journal = "Phys. Rev. D",
    volume = "85",
    pages = "122006",
    year = "2012"
}

@article{Pratten:2020ceb,
    author = "Pratten, Geraint and others",
    title = "{Computationally efficient models for the dominant and subdominant harmonic modes of precessing binary black holes}",
    eprint = "2004.06503",
    archivePrefix = "arXiv",
    primaryClass = "gr-qc",
    doi = "10.1103/PhysRevD.103.104056",
    journal = "Phys. Rev. D",
    volume = "103",
    number = "10",
    pages = "104056",
    year = "2021"
}

@misc{LIGODesign,
    howpublished = {\url{https://dcc.ligo.org/LIGO-T1800044/public}}
}

@article{George:2016hay,
    author = "George, Daniel and Huerta, E. A.",
    title = "{Deep Neural Networks to Enable Real-time Multimessenger Astrophysics}",
    eprint = "1701.00008",
    archivePrefix = "arXiv",
    primaryClass = "astro-ph.IM",
    doi = "10.1103/PhysRevD.97.044039",
    journal = "Phys. Rev. D",
    volume = "97",
    number = "4",
    pages = "044039",
    year = "2018"
}

@article{Cuoco:2024cdk,
    author = "Cuoco, Elena and Cavagli{\`a}, Marco and Heng, Ik Siong and Keitel, David and Messenger, Christopher",
    title = "{Applications of machine learning in gravitational-wave research with current interferometric detectors}",
    eprint = "2412.15046",
    archivePrefix = "arXiv",
    primaryClass = "gr-qc",
    reportNumber = "LIGO-P2400077",
    doi = "10.1007/s41114-024-00055-8",
    journal = "Living Rev. Rel.",
    volume = "28",
    number = "1",
    pages = "2",
    year = "2025"
}

@article{Raikman:2023ktu,
    author = "Raikman, Ryan and others",
    title = "{GWAK: gravitational-wave anomalous knowledge with recurrent autoencoders}",
    eprint = "2309.11537",
    archivePrefix = "arXiv",
    primaryClass = "astro-ph.IM",
    doi = "10.1088/2632-2153/ad3a31",
    journal = "Mach. Learn. Sci. Tech.",
    volume = "5",
    number = "2",
    pages = "025020",
    year = "2024"
}

@article{LIGOScientific:2025slb,
    author = "Abac, A. G. and others",
    collaboration = "LIGO Scientific, VIRGO, KAGRA",
    title = "{GWTC-4.0: Updating the Gravitational-Wave Transient Catalog with Observations from the First Part of the Fourth LIGO-Virgo-KAGRA Observing Run}",
    eprint = "2508.18082",
    archivePrefix = "arXiv",
    primaryClass = "gr-qc",
    reportNumber = "LIGO-P2400386",
    month = "8",
    year = "2025", 
    journal = ""
}

@inproceedings{Kingma:2014vow,
    author = "Kingma, Diederik P. and Ba, Jimmy",
    title = "{Adam: A Method for Stochastic Optimization}",
    eprint = "1412.6980",
    archivePrefix = "arXiv",
    primaryClass = "cs.LG",
    month = "12",
    year = "2014",
    booktitle = ""
}

@article{Harris:2020xlr,
    author = "Harris, Charles R. and others",
    title = "{Array programming with NumPy}",
    eprint = "2006.10256",
    archivePrefix = "arXiv",
    primaryClass = "cs.MS",
    doi = "10.1038/s41586-020-2649-2",
    journal = "Nature",
    volume = "585",
    number = "7825",
    pages = "357--362",
    year = "2020"
}

@article{Virtanen:2019joe,
    author = "Virtanen, Pauli and others",
    title = "{SciPy 1.0--Fundamental Algorithms for Scientific Computing in Python}",
    eprint = "1907.10121",
    archivePrefix = "arXiv",
    primaryClass = "cs.MS",
    doi = "10.1038/s41592-019-0686-2",
    journal = "Nature Meth.",
    volume = "17",
    pages = "261",
    year = "2020"
}

@article{Hunter:2007ouj,
    author = "Hunter, John D.",
    title = "{Matplotlib: A 2D Graphics Environment}",
    doi = "10.1109/MCSE.2007.55",
    journal = "Comput. Sci. Eng.",
    volume = "9",
    number = "3",
    pages = "90--95",
    year = "2007"
}

@article{Astropy:2013muo,
    author = "Robitaille, Thomas P. and others",
    collaboration = "Astropy",
    title = "{Astropy: A Community Python Package for Astronomy}",
    eprint = "1307.6212",
    archivePrefix = "arXiv",
    primaryClass = "astro-ph.IM",
    doi = "10.1051/0004-6361/201322068",
    journal = "Astron. Astrophys.",
    volume = "558",
    pages = "A33",
    year = "2013"
}

@article{Astropy:2018wqo,
    author = "Price-Whelan, A. M. and others",
    collaboration = "Astropy",
    title = "{The Astropy Project: Building an Open-science Project and Status of the v2.0 Core Package}",
    eprint = "1801.02634",
    archivePrefix = "arXiv",
    doi = "10.3847/1538-3881/aabc4f",
    journal = "Astron. J.",
    volume = "156",
    number = "3",
    pages = "123",
    year = "2018"
}

@article{Astropy:2022ucr,
    author = "Price-Whelan, Adrian M. and others",
    collaboration = "Astropy",
    title = "{The Astropy Project: Sustaining and Growing a Community-oriented Open-source Project and the Latest Major Release (v5.0) of the Core Package*}",
    eprint = "2206.14220",
    archivePrefix = "arXiv",
    primaryClass = "astro-ph.IM",
    doi = "10.3847/1538-4357/ac7c74",
    journal = "Astrophys. J.",
    volume = "935",
    number = "2",
    pages = "167",
    year = "2022"
}

@software{alex_nitz_2024_10473621,
  author       = {Alex Nitz and
                  Ian Harry and
                  Duncan Brown and
                  Christopher M. Biwer and
                  Josh Willis and
                  Tito Dal Canton and
                  Collin Capano and
                  Thomas Dent and
                  Larne Pekowsky and
                  Gareth S Cabourn Davies and
                  Soumi De and
                  Miriam Cabero and
                  Shichao Wu and
                  Andrew R. Williamson and
                  Bernd Machenschalk and
                  Duncan Macleod and
                  Francesco Pannarale and
                  Prayush Kumar and
                  Steven Reyes and
                  dfinstad and
                  Sumit Kumar and
                  Márton Tápai and
                  Leo Singer and
                  Praveen Kumar and
                  veronica-villa and
                  maxtrevor and
                  Bhooshan Uday Varsha Gadre and
                  Sebastian Khan and
                  Stephen Fairhurst and
                  Arthur Tolley},
  title        = {gwastro/pycbc: v2.3.3 release of PyCBC},
  month        = jan,
  year         = 2024,
  publisher    = {Zenodo},
  version      = {v2.3.3},
  doi          = {10.5281/zenodo.10473621},
  url          = {https://doi.org/10.5281/zenodo.10473621},
}

@article{Paszke:2019xhz,
    author = "Paszke, Adam and others",
    title = "{PyTorch: An Imperative Style, High-Performance Deep Learning Library}",
    eprint = "1912.01703",
    archivePrefix = "arXiv",
    primaryClass = "cs.LG",
    month = "12",
    year = "2019",
    journal = ""
}

@article{Schafer:2021fea,
    author = {Sch{\"a}fer, Marlin B. and Zelenka, Ond{\v{r}}ej and Nitz, Alexander H. and Ohme, Frank and Br{\"u}gmann, Bernd},
    title = "{Training strategies for deep learning gravitational-wave searches}",
    eprint = "2106.03741",
    archivePrefix = "arXiv",
    primaryClass = "astro-ph.IM",
    doi = "10.1103/PhysRevD.105.043002",
    journal = "Phys. Rev. D",
    volume = "105",
    number = "4",
    pages = "043002",
    year = "2022"
}

@article{Schafer:2022dxv,
    author = {Sch{\"a}fer, Marlin B. and others},
    title = "{First machine learning gravitational-wave search mock data challenge}",
    eprint = "2209.11146",
    archivePrefix = "arXiv",
    primaryClass = "astro-ph.IM",
    doi = "10.1103/PhysRevD.107.023021",
    journal = "Phys. Rev. D",
    volume = "107",
    number = "2",
    pages = "023021",
    year = "2023"
}

@software{marlin_schafer_2022_7107410,
  author       = {Marlin Schäfer and
                  Ondřej Zelenka and
                  Pascal Müller and
                  Alex Nitz},
  title        = {gwastro/ml-mock-data-challenge-1: MLGWSC-1 Release
                   v1.4
                  },
  month        = sep,
  year         = 2022,
  publisher    = {Zenodo},
  version      = {v1.4},
  doi          = {10.5281/zenodo.7107410},
  url          = {https://doi.org/10.5281/zenodo.7107410},
}

@article{Joshi:2023hpx,
    author = "Joshi, Prasanna M. and Prix, Reinhard",
    title = "{Novel neural-network architecture for continuous gravitational waves}",
    eprint = "2305.01057",
    archivePrefix = "arXiv",
    primaryClass = "gr-qc",
    doi = "10.1103/PhysRevD.108.063021",
    journal = "Phys. Rev. D",
    volume = "108",
    number = "6",
    pages = "063021",
    year = "2023"
}

@article{Joshi:2024bkj,
    author = "Joshi, Prasanna Mohan and Prix, Reinhard",
    title = "{Large-kernel convolutional neural networks for wide parameter-space searches of continuous gravitational waves}",
    eprint = "2408.07070",
    archivePrefix = "arXiv",
    primaryClass = "gr-qc",
    doi = "10.1103/PhysRevD.110.124071",
    journal = "Phys. Rev. D",
    volume = "110",
    number = "12",
    pages = "124071",
    year = "2024"
}

@article{Joshi:2025xdz,
    author = "Joshi, Prasanna. M. and Prix, Reinhard",
    title = "{Transformer Networks for Continuous Gravitational-wave Searches}",
    eprint = "2509.10912",
    archivePrefix = "arXiv",
    primaryClass = "gr-qc",
    month = "9",
    year = "2025",
    journal = ""
}

@article{Yamamoto:2020pus,
    author = "Yamamoto, Takahiro S. and Tanaka, Takahiro",
    title = "{Use of an excess power method and a convolutional neural network in an all-sky search for continuous gravitational waves}",
    eprint = "2011.12522",
    archivePrefix = "arXiv",
    primaryClass = "gr-qc",
    doi = "10.1103/PhysRevD.103.084049",
    journal = "Phys. Rev. D",
    volume = "103",
    number = "8",
    pages = "084049",
    year = "2021"
}

@article{Cheung:2025lyf,
    author = "Cheung, Damon H. T.",
    title = "{Attention U-Net for all-sky continuous gravitational wave searches}",
    eprint = "2509.19838",
    archivePrefix = "arXiv",
    primaryClass = "gr-qc",
    doi = "10.1103/6spt-23vj",
    journal = "Phys. Rev. D",
    volume = "112",
    number = "10",
    pages = "103039",
    year = "2025"
}

@article{Dreissigacker:2019edy,
    author = "Dreissigacker, Christoph and Sharma, Rahul and Messenger, Chris and Zhao, Ruining and Prix, Reinhard",
    title = "{Deep-Learning Continuous Gravitational Waves}",
    eprint = "1904.13291",
    archivePrefix = "arXiv",
    primaryClass = "gr-qc",
    doi = "10.1103/PhysRevD.100.044009",
    journal = "Phys. Rev. D",
    volume = "100",
    number = "4",
    pages = "044009",
    year = "2019"
}

@article{Dreissigacker:2020xfr,
    author = "Dreissigacker, Christoph and Prix, Reinhard",
    title = "{Deep-Learning Continuous Gravitational Waves: Multiple detectors and realistic noise}",
    eprint = "2005.04140",
    archivePrefix = "arXiv",
    primaryClass = "gr-qc",
    doi = "10.1103/PhysRevD.102.022005",
    journal = "Phys. Rev. D",
    volume = "102",
    number = "2",
    pages = "022005",
    year = "2020"
}

@article{Yamamoto:2022adl,
    author = "Yamamoto, Takahiro S. and Miller, Andrew L. and Sieniawska, Magdalena and Tanaka, Takahiro",
    title = "{Assessing the impact of non-Gaussian noise on convolutional neural networks that search for continuous gravitational waves}",
    eprint = "2206.00882",
    archivePrefix = "arXiv",
    primaryClass = "gr-qc",
    doi = "10.1103/PhysRevD.106.024025",
    journal = "Phys. Rev. D",
    volume = "106",
    number = "2",
    pages = "024025",
    year = "2022"
}

@article{Einsle:2025xsh,
    author = "Einsle, Hugo and Bizouard, Marie-Anne and Regimbau, Tania and Sakellariadou, Mairi",
    title = "{Gravitational-wave background detection using machine learning}",
    eprint = "2506.14764",
    archivePrefix = "arXiv",
    primaryClass = "gr-qc",
    doi = "10.1103/hs9b-drwx",
    journal = "Phys. Rev. D",
    volume = "112",
    number = "6",
    pages = "063056",
    year = "2025"
}

@article{Yamamoto:2022kuh,
    author = "Yamamoto, Takahiro S. and Kuroyanagi, Sachiko and Liu, Guo-Chin",
    title = "{Deep learning for intermittent gravitational wave signals}",
    eprint = "2208.13156",
    archivePrefix = "arXiv",
    primaryClass = "gr-qc",
    doi = "10.1103/PhysRevD.107.044032",
    journal = "Phys. Rev. D",
    volume = "107",
    number = "4",
    pages = "044032",
    year = "2023"
}

@article{Astone:2018uge,
    author = "Astone, P. and Cerd{\'a}-Dur{\'a}n, P. and Di Palma, I. and Drago, M. and Muciaccia, F. and Palomba, C. and Ricci, F.",
    title = "{New method to observe gravitational waves emitted by core collapse supernovae}",
    eprint = "1812.05363",
    archivePrefix = "arXiv",
    primaryClass = "astro-ph.IM",
    doi = "10.1103/PhysRevD.98.122002",
    journal = "Phys. Rev. D",
    volume = "98",
    number = "12",
    pages = "122002",
    year = "2018"
}

@article{Iess:2020yqj,
    author = "Iess, Alberto and Cuoco, Elena and Morawski, Filip and Powell, Jade",
    title = "{Core-Collapse Supernova Gravitational-Wave Search and Deep Learning Classification}",
    eprint = "2001.00279",
    archivePrefix = "arXiv",
    primaryClass = "gr-qc",
    month = "1",
    year = "2020",
    journal = ""
}

@article{Iess:2023quq,
    author = "Iess, Alberto and Cuoco, Elena and Morawski, Filip and Nicolaou, Constantina and Lahav, Ofer",
    title = "{LSTM and CNN application for core-collapse supernova search in gravitational wave real data}",
    eprint = "2301.09387",
    archivePrefix = "arXiv",
    primaryClass = "astro-ph.IM",
    doi = "10.1051/0004-6361/202142525",
    journal = "Astron. Astrophys.",
    volume = "669",
    pages = "A42",
    year = "2023"
}

@article{Chan:2019fuz,
    author = "Chan, Man Leong and Heng, Ik Siong and Messenger, Chris",
    title = "{Detection and classification of supernova gravitational wave signals: A deep learning approach}",
    eprint = "1912.13517",
    archivePrefix = "arXiv",
    primaryClass = "astro-ph.HE",
    doi = "10.1103/PhysRevD.102.043022",
    journal = "Phys. Rev. D",
    volume = "102",
    number = "4",
    pages = "043022",
    year = "2020"
}

@article{Edwards:2020hmd,
    author = "Edwards, Matthew C.",
    title = "{Classifying the Equation of State from Rotating Core Collapse Gravitational Waves with Deep Learning}",
    eprint = "2009.07367",
    archivePrefix = "arXiv",
    primaryClass = "astro-ph.IM",
    doi = "10.1103/PhysRevD.103.024025",
    journal = "Phys. Rev. D",
    volume = "103",
    number = "2",
    pages = "024025",
    year = "2021"
}

@article{Abylkairov:2024hjf,
    author = "Abylkairov, Y. Sultan and Edwards, Matthew C. and Orel, Daniil and Mitra, Ayan and Shukirgaliyev, Bekdaulet and Abdikamalov, Ernazar",
    title = "{Evaluating machine learning models for supernova gravitational wave signal classification}",
    eprint = "2409.14508",
    archivePrefix = "arXiv",
    primaryClass = "astro-ph.HE",
    doi = "10.1088/2632-2153/ada33a",
    journal = "Mach. Learn. Sci. Tech.",
    volume = "5",
    number = "4",
    pages = "045077",
    year = "2024"
}

@article{LopezPortilla:2020odz,
    author = "L{\'o}pez Portilla, M. and Palma, I. Di and Drago, M. and Cerd{\'a}-Dur{\'a}n, P. and Ricci, F.",
    title = "{Deep learning for core-collapse supernova detection}",
    eprint = "2011.13733",
    archivePrefix = "arXiv",
    primaryClass = "astro-ph.IM",
    doi = "10.1103/PhysRevD.103.063011",
    journal = "Phys. Rev. D",
    volume = "103",
    number = "6",
    pages = "063011",
    year = "2021"
}

@article{Sasaoka:2023kte,
    author = "Sasaoka, Seiya and Koyama, Naoki and Dominguez, Diego and Sakai, Yusuke and Somiya, Kentaro and Omae, Yuto and Takahashi, Hirotaka",
    title = "{Visualizing convolutional neural network for classifying gravitational waves from core-collapse supernovae}",
    eprint = "2310.09551",
    archivePrefix = "arXiv",
    primaryClass = "astro-ph.IM",
    doi = "10.1103/PhysRevD.108.123033",
    journal = "Phys. Rev. D",
    volume = "108",
    number = "12",
    pages = "123033",
    year = "2023"
}

@article{Sakai:2021rks,
    author = "Sakai, Yusuke and others",
    title = "{Unsupervised learning architecture for classifying the transient noise of interferometric gravitational-wave detectors}",
    eprint = "2111.10053",
    archivePrefix = "arXiv",
    primaryClass = "gr-qc",
    reportNumber = "JGW-P2113444",
    doi = "10.1038/s41598-022-13329-4",
    journal = "Sci. Rep.",
    volume = "12",
    number = "1",
    pages = "9935",
    year = "2022"
}

@article{Nousi:2022dwh,
    author = "Nousi, Paraskevi and Koloniari, Alexandra E. and Passalis, Nikolaos and Iosif, Panagiotis and Stergioulas, Nikolaos and Tefas, Anastasios",
    title = "{Deep residual networks for gravitational wave detection}",
    eprint = "2211.01520",
    archivePrefix = "arXiv",
    primaryClass = "gr-qc",
    doi = "10.1103/PhysRevD.108.024022",
    journal = "Phys. Rev. D",
    volume = "108",
    number = "2",
    pages = "024022",
    year = "2023"
}

@article{Koloniari:2024kww,
    author = "Koloniari, Alexandra E. and Koursoumpa, Evdokia C. and Nousi, Paraskevi and Lampropoulos, Paraskevas and Passalis, Nikolaos and Tefas, Anastasios and Stergioulas, Nikolaos",
    title = "{New gravitational wave discoveries enabled by machine learning}",
    eprint = "2407.07820",
    archivePrefix = "arXiv",
    primaryClass = "gr-qc",
    doi = "10.1088/2632-2153/adb5ed",
    journal = "Mach. Learn. Sci. Tech.",
    volume = "6",
    number = "1",
    pages = "015054",
    year = "2025"
}

@article{Koloniari:2025cyt,
    author = "Koloniari, Alexandra E. and Lazaridis, Lazaros and Paschalidis, Christos and Stergioulas, Nikolaos",
    title = "{Robustness of Sensitivity Evaluations for Gravitational Wave Detection Algorithms}",
    eprint = "2509.05283",
    archivePrefix = "arXiv",
    primaryClass = "gr-qc",
    month = "9",
    year = "2025",
    journal= ""
}

@article{Nagarajan:2025hws,
    author = "Nagarajan, Narenraju and Messenger, Christopher",
    title = "{Identifying and mitigating machine-learning biases for the gravitational-wave detection problem}",
    eprint = "2501.13846",
    archivePrefix = "arXiv",
    primaryClass = "gr-qc",
    doi = "10.1103/zwj9-ycyz",
    journal = "Phys. Rev. D",
    volume = "112",
    number = "10",
    pages = "103002",
    year = "2025"
}

@article{Wang:2019zaj,
    author = "Wang, He and Wu, Shichao and Cao, Zhoujian and Liu, Xiaolin and Zhu, Jian-Yang",
    title = "{Gravitational-wave signal recognition of LIGO data by deep learning}",
    eprint = "1909.13442",
    archivePrefix = "arXiv",
    primaryClass = "astro-ph.IM",
    doi = "10.1103/PhysRevD.101.104003",
    journal = "Phys. Rev. D",
    volume = "101",
    number = "10",
    pages = "104003",
    year = "2020"
}

@article{Gabbard:2017lja,
    author = "Gabbard, Hunter and Williams, Michael and Hayes, Fergus and Messenger, Chris",
    title = "{Matching matched filtering with deep networks for gravitational-wave astronomy}",
    eprint = "1712.06041",
    archivePrefix = "arXiv",
    primaryClass = "astro-ph.IM",
    doi = "10.1103/PhysRevLett.120.141103",
    journal = "Phys. Rev. Lett.",
    volume = "120",
    number = "14",
    pages = "141103",
    year = "2018"
}

@article{Schafer:2021cml,
    author = {Sch{\"a}fer, Marlin B. and Nitz, Alexander H.},
    title = "{From one to many: A deep learning coincident gravitational-wave search}",
    eprint = "2108.10715",
    archivePrefix = "arXiv",
    primaryClass = "astro-ph.IM",
    doi = "10.1103/PhysRevD.105.043003",
    journal = "Phys. Rev. D",
    volume = "105",
    number = "4",
    pages = "043003",
    year = "2022"
}
\end{document}